\documentclass[usenatbib]{mn2e}

\usepackage[dvips]{epsfig,color}

\usepackage{amssymb,amsmath}

\newcommand{\eg}{e.g.}
\newcommand{\ie}{i.e.}
\newcommand{\etal}{et al.}

\newcommand{\ignore}[1]{\relax}

\DeclareMathAlphabet{\mathsfsl}{OT1}{cmss}{m}{sl}
\DeclareMathOperator{\sech}{sech}

\newcommand{\dif}{\mathrm{d}}
\newcommand{\avgv}{\ensuremath{\langle \varpi \rangle}}
\newcommand{\avgvv}{\ensuremath{\langle \varpi^2 \rangle}}
\newcommand{\avgvvv}{\ensuremath{\langle \varpi^3 \rangle}}

\begin{document}

\title{Relaxation of One-dimensional Collisionless Gravitating
Systems}
\author[Barnes \& Ragan]{
Eric I. Barnes\thanks{email:barnes.eric@uwlax.edu}, Robert J.
Ragan\thanks{email:rragan@uwlax.edu} \\
Department of Physics, University of Wisconsin --- La
Crosse, La Crosse, WI 54601}

\maketitle

\begin{abstract}

In an effort to better understand collisionless relaxation processes
in gravitational systems, we investigate one-dimensional models.
Taking advantage of a Hermite-Legendre expansion of relevant
distribution functions, we present analytical and numerical behaviors
of Maxwell-Boltzmann entropy.  In particular, we modestly perturb
systems about a separable-solution equilibrium and observe their
collisionless evolution to a steady state.  We verify the
time-independence of fine-grained entropy in these systems before
turning our attention to the behavior of coarse-grained entropy.  We
also verify that there is no analogue to the collisional H-theorem for
these systems.  Competing terms in the second-order coarse-grained
entropy make it impossible to guarantee continuously increasing
entropy.  However, over dynamical time-scales the coarse-grained
entropy generally increases, with small oscillations occurring. The
lack of substantive differences between the entropies in test-particle
and self-gravitating cases suggests that phase mixing, rather than
violent relaxation associated with potential changes, more
significantly drives the coarse-grained entropy evolution.  The
effects of violent relaxation can be better quantified through
analysis of energy distributions rather than phase-space
distributions.

\end{abstract}


\section{Introduction}\label{intro}

Thermodynamics of self-gravitating collisionless systems is an
interesting subject.  In three-dimensions, such systems do
not have thermodynamic equilibria, characterized by a constant kinetic
temperature or a maximum entropy \citep[\eg][\S 4.7]{bt87}.
Investigations of statistical mechanics and thermodynamic approaches
to understanding the mechanical equilibria of self-gravitating systems
have a long history.  A succinct introduction to the problem and
historical review of progress made is given by \citet{letal14}.  

In this work, we take on a related, but simpler, situation.
One-dimensional self-gravitating systems can be thought of either as
particles interacting in one-dimension through a distance-independent
force or as infinite sheets of mass in three-dimensions.
Collisionless versions of these systems do have equilibria with
simple, separable forms \citep{c50}.  In fact, the separable
equilibrium distribution function has Boltzmann form.  Readers
interested in the nature of this equilibrium and its relationship to
statistical equilibria in the microcanonical and canonical ensembles
may find previous work by \citet{r71} and \citet{j10} particularly
interesting.  The temperature of such an equilibrium can be connected
to an energy scaling factor, as in normal collisional gas situations.
Additionally, the kinetic and thermal temperatures $kT=1/\beta$ are
identical for these equilibria.

Our goal is to understand collisionless relaxation from an
initial state that is perturbed from this equilibrium, which we pursue
by investigating the behavior of Maxwell-Boltzmann entropy.  Based on
the work of \citet*{tetal86}, we use the Maxwell-Boltzmann qualifier
here to specify the common form of the entropy function we will deal
with.  On a side note, we have also confirmed that the Lynden-Bell
entropy \citep{lb67,bw12} behaves essentially identically to the
Maxwell-Boltzmann entropy in these non-degenerate systems.  For
simplicity, we will assume the reader implicitly inserts the
Maxwell-Boltzmann qualifier to all further references to entropy.  We
investigate how well the Tremaine \etal\ analogy to the collisional
H-theorem (guaranteeing that entropy increases during relaxation)
explains aspects of the dynamics of these systems.  We also
compare the evolutions of systems composed of test particles to
self-gravitating situations in an attempt to separate the influences
of phase mixing and violent relaxation.

Phase mixing describes evolution in which any occupied region of phase
space tends to mix with unoccupied phase space, producing a lower
average phase-space density.  Particle interactions are not necessary
for phase mixing to occur.  Violent relaxation, on the other hand,
generally results when the dynamics of a system are driven by a
time-dependent potential \citep{lb67}.  In our simulations, violent
relaxation is driven by self-gravitation in systems that are in
non-stationary states.  In general, phase mixing occurs in
all of our simulated systems, test-particle and self-gravitating, but
violent relaxation is absent from test-particle situations.

While we do use $N$-body simulations, the basis of our analysis takes
advantage of a Hermite-Legendre decomposition of distribution
functions.  This effectively changes the problem from one involving
continuous phase-space coordinates to a discrete coefficient space
situation.  The dynamics of an evolving system with infinite range
forces then reduces to a local interaction between coefficients.  The
basics of this approach are presented in \citet{br14}.  We will expand
upon the arbitrary perturbation discussion in \citet{rb19a} when
dealing with second-order effects in what follows.

The introduction to essential features of our approach (coefficient
dynamics and second-order perturbation theory) is provided in the
discussion of energy given in \S~\ref{esec}.  The entropy analysis for
both fine- and coarse-grained situations follows in
Section~\ref{entsec}.  We also highlight agreement between our
coefficient results and a thermodynamical approach to calculating
entropy changes.  The importance of energy distributions, as opposed
to entropy, in quantifying violent relaxation is explored in
Section~\ref{nofe}. Our approach and results are summarized in
\S~\ref{sum}.

\section{Energy}\label{esec}

\subsection{General Relationships}

For a one-dimensional situation, phase space is simply a
position-velocity plane $(x,v)$.  We adopt dimensionless versions of
position and velocity using system mass $M$, gravitational coupling
constant $g$, and an energy scale $\beta$.  Specifically,
\begin{displaymath}
\chi=\frac{\beta gM}{2} x \quad \mbox{and} \quad
\varpi=\sqrt{\frac{\beta}{2}}v.
\end{displaymath}
The dimensionless time is then defined by
\begin{displaymath}
\tau=\sqrt{\frac{\beta}{2}}gMt.
\end{displaymath}
We adopt these dimensionless coordinates for the
remainder of this paper.

For any distribution function $f(\chi,\varpi)$, we define the
dimensionless mass density function as,
\begin{equation}\label{density}
\Lambda = \int_{-\infty}^{\infty} f \, \dif \varpi.
\end{equation}
We also define the dimensionless one-dimensional self-gravitating
acceleration as,
\begin{equation}\label{accel}
\alpha = \frac{a}{gM} = -\int_{-\infty}^{\chi} \Lambda(\chi^{\prime})
\, \dif \chi^{\prime} + \int_{\chi}^{\infty} \Lambda(\chi^{\prime}) \,
\dif \chi^{\prime}.
\end{equation}
Consider a distribution function decomposition in terms of Hermite and
Legendre polynomials,
\begin{equation}\label{fexpand}
f(\chi^{\prime \prime},\varpi) = \sum_{m,n=0}^{\infty} C_{m,n}
H_m(\varpi) P_n(\chi^{\prime \prime}) \sech^2 \chi^{\prime \prime}
e^{-\varpi^2}.
\end{equation}
We note that the coefficients in this decomposition are modified from
those in \citet{br14} and \citet{rb19a}.  The coefficients here have
absorbed factors of $\sqrt{(2n+1)/2}$ and $1/\sqrt{2^m \sqrt{\pi}
m!}$.  With this choice we have that,
\begin{equation}\label{dspec}
\Lambda = \sqrt{\pi} \sum_{n=0}^{\infty} 
C_{0,n} P_n \sech^2 \chi^{\prime \prime},
\end{equation}
where we have taken advantage of the orthogonality relations for
Hermite polynomials,
\begin{displaymath}
\int_{-\infty}^{\infty} H_j(\varpi) H_k(\varpi) \, \dif \varpi = 2^j
\sqrt{\pi} j! \, \delta_{jk},
\end{displaymath}
where $\delta$ is the Kronecker delta.  With this density
(Equation~\ref{dspec}), the acceleration becomes,
\begin{equation}
\alpha = \sqrt{\pi} \sum_{n=0}^{\infty} C_{0,n} 
\left[ \int_{\tanh \chi^{\prime}}^1 P_n(u) \, \dif u - 
\int_{-1}^{\tanh \chi^{\prime}} P_n(u) \, \dif u \right],
\end{equation}
where we have made the substitution $u=\tanh \chi$.  These integrals
may be carried out to produce an expression for acceleration purely in
terms of Legendre polynomials,
\begin{equation}
\alpha = -2\sqrt{\pi} \sum_{n \ge 0} \frac{C_{0,n}}{2n+1} \left[
P_{n+1}(\tanh \chi^{\prime}) - P_{n-1}(\tanh \chi^{\prime}) \right].
\end{equation}
The dimensionless potential function may now be written as,
\begin{eqnarray}
\Phi(\chi) - \Phi(0) & = & 4\sqrt{\pi} \sum_{n \ge 0} \frac{C_{0,n}}{2n+1} 
\left[ \int_0^{\chi} P_{n+1}(\tanh \chi^{\prime}) \, \dif
\chi^{\prime} - \right. \nonumber \\
 & & \left. \int_0^{\chi} P_{n-1}(\tanh \chi^{\prime}) \, \dif
\chi^{\prime} \right].
\end{eqnarray}
Again taking advantage of the substitution $u=\tanh \chi$, the
integrals may be combined into,
\begin{equation}\label{phiint}
\int_0^{\tanh \chi} \frac{P_{n+1}(u)-P_{n-1}(u)}{1-u^2} \, \dif u.
\end{equation}
Using that
\begin{displaymath}
\frac{\dif P_N}{\dif u} = \frac{N}{1-u^2} (P_{N-1}-uP_N),
\end{displaymath}
and
\begin{displaymath}
uP_N = \frac{N+1}{2N+1}P_{N+1}+\frac{N}{2N+1} P_{N-1},
\end{displaymath}
one can show that Equation~\ref{phiint} reduces to
\begin{equation}
-\frac{2n+1}{n(n+1)} \left[ P_n(\tanh \chi) - P_n(0) \right],
\end{equation}
for $n>0$.  When $n=0$, Equation~\ref{phiint} is simply $\ln{(\cosh
\chi)}$.  Using the fact that $P_n(0)$ is zero for odd $n$ and
demanding that $\lim_{\chi \rightarrow \infty} \Phi = 2\chi$ (system
mass has finite extent) leads to,
\begin{equation}\label{phi}
\Phi(\chi) = 2 \ln{(2 \cosh \chi)} + 4\sqrt{\pi} \sum_{n\ge 1}
\frac{C_{0,n}}{n(n+1)} \left[ 1- P_n(\tanh \chi) \right].
\end{equation}

With this expression and Equation~\ref{fexpand}, the dimensionless
potential energy of the system may be written as,
\begin{eqnarray}
U & = &
\frac{1}{2}\int_{-\infty}^{\infty} \int_{-\infty}^{\infty} f \Phi
\, \dif \chi \dif \varpi \nonumber \\
& = & 1 + 2\sqrt{\pi} \sum_{n \ge 1\atop {\rm odd}}
\frac{C_{0,n}}{n(n+1)} + 4\sqrt{\pi} \sum_{n \ge 2\atop {\rm even}}
\frac{C_{0,n}}{n(n+1)} - \nonumber \\ & & 4\pi \sum_{n \ge 1}
\frac{C_{0,n}^2}{n(n+1)(2n+1)}.
\end{eqnarray}
Given that the dimensionless kinetic energy can be expressed as
\citep{br14},
\begin{eqnarray}
K & = & \int_{-\infty}^{\infty}
\int_{-\infty}^{\infty} f \varpi^2 \, \dif \chi \dif \varpi
\nonumber \\
& = & \sqrt{\pi} (C_{0,0} + 4 C_{2,0}) \nonumber \\
& = & \frac{1}{2} + 4\sqrt{\pi} C_{2,0},
\end{eqnarray}
the Hermite-Legendre decomposition provides an interesting picture of
energy behavior in these systems in ($m,n$) coefficient space.  In
linear perturbation regimes, coefficient dynamics equations demand
that only diagonally neighboring coefficients can interact
\citep{br14}.  In this way, kinetic energy changes always link
directly to large-spatial-scale variations in the potential ($m=0,n=2$
in Equation~\ref{phi}) which then propagate to smaller-spatial-scale
potential variations with larger $n$.  This energy flow to higher $n$
coefficients passes through $m=1$ terms, which have to act as
conduits, as they cannot directly contribute to the total energy.

\subsection{Perturbation Analysis}\label{pertsec}

As mentioned above, the dynamics of the system can be written in terms
of decomposition coefficients if the perturbation strength is kept
small.  The original coefficient evolution investigation in
\citet{br14} only included first-order terms.  However, from the
general potential energy expression above, one can see that complete
potential energy calculations require second-order terms.  As we are
interested in capturing violent relaxation processes involving
potential changes, we need to extend our perturbation analysis to
second-order as well.

This leads us to consider systems such that the distribution function
can be written as,
\begin{equation}\label{pertf}
f = f_0 + \epsilon f_1 + \epsilon^2 f_2,
\end{equation}
where 
\begin{displaymath}
f_0=\frac{1}{2\sqrt{\pi}} \sech^2 \chi e^{-\varpi^2} 
\end{displaymath}
describes the separable-solution equilibrium (hereafter,
separable equilibrium for brevity) and $\epsilon^2 \ll
1$.  The perturbation functions have similar forms,
\begin{equation}\label{ord1}
f_1=\sum_{m,n\atop m\ne n=0} c_{m,n} H_m(\varpi) P_n(\tanh \chi)
\sech^2 \chi e^{-\varpi^2},
\end{equation}
and
\begin{equation}\label{ord2}
f_2=\sum_{m,n\atop m\ne n=0} d_{m,n} H_m(\varpi) P_n(\tanh \chi)
\sech^2 \chi e^{-\varpi^2},
\end{equation}
where the $m=n=0$ terms are explicitly excluded.

Using the results of the previous section, we calculate the
dimensionless energy of our system as,
\begin{equation}
E = K + U = \int_{-\infty}^{\infty}
\int_{-\infty}^{\infty} f \varpi^2 \, \dif \chi \dif \varpi +
\frac{1}{2}\int_{-\infty}^{\infty} \int_{-\infty}^{\infty} f \Phi
\, \dif \chi \dif \varpi.
\end{equation}
To second order, the kinetic energy term is,
\begin{eqnarray}
K & = & K_0 + \epsilon K_1 + \epsilon^2 K_2 \nonumber \\ 
& = &
\iint f_0 \varpi^2 \, \dif \chi \dif \varpi + \epsilon \iint f_1
\varpi^2 \, \dif \chi \dif \varpi + \nonumber \\ & & \epsilon^2 \iint
f_2 \varpi^2 \, \dif \chi \dif \varpi.
\end{eqnarray}
From here on, infinite limits of integration should be assumed
wherever limits are omitted.  Completing these integrations produces,
\begin{eqnarray}
K_0 & = & \frac{1}{2}, \\
K_1 & = & 4\sqrt{\pi} c_{2,0}, \quad \mbox{and} \nonumber \\
K_2 & = & 4\sqrt{\pi} d_{2,0}. \nonumber
\end{eqnarray}
The potential energy contribution is,
\begin{eqnarray}\label{pote2int}
U & = & U_0 + \epsilon U_1 + \epsilon^2 U_2 = \nonumber \\ & &
\frac{1}{2} \iint f_0 \Phi_0 \, \dif \chi \dif \varpi +
\frac{\epsilon}{2} \iint (f_1 \Phi_0 + f_0 \Phi_1) \, \dif \chi
\dif \varpi + \nonumber \\ & & \frac{\epsilon^2}{2} \iint (f_2
\Phi_0 + f_1 \Phi_1 + f_0 \Phi_2) \, \dif \chi \dif \varpi,
\end{eqnarray}
where the infinite limits of integration are assumed.
Here, $\Phi_0=2\ln{(2 \cosh \chi)}$ and,
\begin{displaymath}
\Phi_1 = 4\sqrt{\pi} \sum_{n\ge 1} \frac{c_{0,n}}{n(n+1)} \left[1 -
P_n(\tanh \chi) \right],
\end{displaymath}
and
\begin{displaymath}
\Phi_2 = 4\sqrt{\pi} \sum_{n\ge 1} \frac{d_{0,n}}{n(n+1)} \left[1 -
P_n(\tanh \chi) \right].
\end{displaymath}

We now write the energy of the system as,
\begin{equation}
E = E_0 + \epsilon E_1 + \epsilon^2 E_2,
\end{equation}
where
\begin{eqnarray}\label{energy}
E_0 & = & \frac{3}{2}, \\
E_1 & = & 4\sqrt{\pi}\left[ c_{2,0} + \sum_{n\ge 2\atop {\rm even}}
\frac{c_{0,n}}{n(n+1)} + \frac{1}{2}\sum_{n\ge 1\atop {\rm odd}}
\frac{c_{0,n}}{n(n+1)} \right], \, \mbox{and} \nonumber \\
E_2 & = & 4\sqrt{\pi}\left[ d_{2,0} + \sum_{n\ge 2\atop {\rm even}}
\frac{d_{0,n}}{n(n+1)} + \frac{1}{2}\sum_{n\ge 1\atop {\rm odd}}
\frac{d_{0,n}}{n(n+1)}\right]- \nonumber \\
& & 4\pi \sum_{n\ge 1} \frac{c_{0,n}^2}{n(n+1)(2n+1)}.\nonumber
\end{eqnarray}
The terms in the square brackets in each of these expressions reflect
the interaction of the perturbation with the equilibrium, but the term
proportional to $c_{0,n}^2$ in the second-order energy is due to the
perturbation interacting with itself.  Its origin lies with the $f_1
\Phi_1$ term in Equation~\ref{pote2int}.

Energy must be conserved in these systems, but in order to prove this
we need dynamics equations for the first- and second-order
coefficients $\dot{c}_{m,n}$ and $\dot{d}_{m,n}$.  The details of the
derivation of the first-order coefficient equation have been presented
in \citet{br14}.  The second-order coefficient equation closely
follows that discussion, but there is an interesting addition.  We
find that,
\begin{eqnarray}\label{cbelp}
\lefteqn{\dot{d}_{m,n} = } \nonumber \\
 & & L_{m,n}^{m-1,n-1} \, d_{m-1,n-1} + 
 L_{m,n}^{m-1,n+1} \, d_{m-1,n+1} + \\ 
& & L_{m,n}^{m+1,n-1}\, d_{m+1,n-1} + L_{m,n}^{m+1,n+1} \, d_{m+1,n+1}
- R_{m,n}, \nonumber
\end{eqnarray}
where the matrix elements $L_{m,n}^{i,j}$ are given by 
\begin{eqnarray}\label{Ldef}
L_{m,n}^{m-1,n-1}&=&\frac{n(n-1)-2\delta_{1,m}}
{2(2n-1)},\nonumber \\
L_{m,n}^{m-1,n+1}&=&-\frac{(n+1)(n+2)-2\delta_{1,m}}
{2(2n+3)},\nonumber \\
L_{m,n}^{m+1,n-1}&=&\frac{(m+1)n(n+1)}
{2n-1},\nonumber \\ 
L_{m,n}^{m+1,n+1}&=&- \frac{(m+1)n(n+1)}{2n+3},
\end{eqnarray}
where $m,n,i,j \ge 0$.  The Kronecker delta functions are present for
self-gravitating systems only.  Test-particle systems do not require
them to determine their dynamics.  Replacing $d$ with $c$ and setting
$R_{m,n}=0$ in Equation~\ref{cbelp} provides the coefficient dynamics
for the first-order perturbation.  The additional term for the
second-order is given by,
\begin{eqnarray}
R_{m,n} & = & 2\sqrt{\pi}(2n+1)\sum_{p\ge 1} \frac{c_{0,p}}{2p+1} 
\left\{ c_{m-1,0} Q_{n+p+1}^{(n,p+1)} + \right. \nonumber \\
& & \sum_{s=0\atop {\rm even}}^{n+p-1} \left[ Q_s^{(n,p+1)}
\left( \frac{c_{m-1,n+p+1-s}}{2(n+p+1-s)+1} \right) - \right. \nonumber \\
& & \left. \left. Q_s^{(n,p-1)} \left( \frac{c_{m-1,n+p-1-s}}{2(n+p-1-s)+1} 
\right) \right] \right\}.
\end{eqnarray}
The $Q$ functions arise from writing products of Legendre polynomials
as series of single Legendre polynomials and are defined by \citep{d53},
\begin{displaymath}
Q_s^{(j,k)} = \frac{2j+2k-2s+1}{2j+2k-s+1} \frac{\lambda_{s/2}
\lambda_{j-s/2} \lambda_{k-s/2}}{\lambda_{j+k-s/2}},
\end{displaymath}
where
\begin{displaymath}
\lambda_B = \frac{(2B)!}{2^B (B!)^2},
\end{displaymath}
if $B \ge 0$ and is zero otherwise.

We will focus on a discussion of the time-derivative of $E_2$.  In
this discussion, we will assume that all even-$m$, odd-$n$
coefficients are zero.  This guarantees that the system center-of-mass
position and velocity are constants.  From Equation~\ref{energy},
\begin{eqnarray}\label{de0}
\dot{E}_2 & = & 4\sqrt{\pi}\left[ \dot{d}_{2,0} + 
\sum_{n\ge 2\atop {\rm even}} \frac{\dot{d}_{0,n}}{n(n+1)} \right]- 
\nonumber \\
& & 4\pi \frac{\partial}{\partial \tau} \left[ \sum_{n\ge 1} 
\frac{c_{0,n}^2}{n(n+1)(2n+1)}\right].
\end{eqnarray}
Using Equation~\ref{cbelp} and the specific values of $Q$, one can
show that,
\begin{equation}\label{de1}
\dot{d}_{2,0} + \sum_{n\ge 2\atop {\rm even}}
\frac{\dot{d}_{0,n}}{n(n+1)}=
-2\sqrt{\pi} \sum_{n\ge 1} \frac{c_{0,n}}{2n+1} \left[
\frac{c_{1,n+1}}{2n+3} - \frac{c_{1,n-1}}{2n-1} \right].
\end{equation}
The term in square brackets on the right-hand side of this expression
can be re-cast using the first-order version of Equation~\ref{cbelp}
(with $R_{m,n}=0$).  We have that,
\begin{equation}
\frac{\dot{c}_{0,n}}{n(n+1)} =
\frac{c_{1,n-1}}{2n-1}-\frac{c_{1,n+1}}{2n+3}.
\end{equation}
Substituting this relation into Equation~\ref{de1} results in,
\begin{equation}\label{de2}
\dot{d}_{2,0} + \sum_{n\ge 2\atop {\rm even}}
\frac{\dot{d}_{0,n}}{n(n+1)}=
\sqrt{\pi} \frac{\partial}{\partial \tau} \left[ \sum_{n\ge 1} 
\frac{c_{0,n}^2}{n(n+1)(2n+1)} \right].
\end{equation}
Using this expression in Equation~\ref{de0} provides us with the proof
that the second-order energy is time-independent.  The proof for the
first-order energy is simpler, as the first-order version
of Equation~\ref{de0} does not include the last term.  It is
then straightforward to show from the first-order coefficient dynamics
equations that,
\begin{equation}\label{cd1}
\dot{c}_{2,0} + \sum_{n\ge 2\atop {\rm even}}
\frac{\dot{c}_{0,n}}{n(n+1)} = 0.
\end{equation}

\section{Entropy}\label{entsec}

In a collisional system, the H-theorem guarantees that entropy
increases as a system approaches equilibrium.  For collisionless
systems, \citet{tetal86} have shown that any convex function of the
distribution function, like the Maxwell-Boltzmann entropy, will not
decrease during relaxation.  In our discussion, we are careful to
distinguish between entropy based on the fine-grained distribution
function and entropy based on a coarse-grained distribution function.
In a collisionless system, the fine-grained entropy is a conserved
quantity, like energy.  A coarse-grained entropy does not have to be
conserved, and we are interested in how its evolution compares to the
Tremaine \etal\ collisionless H-theorem prediction.  Our goal in this
section is to derive and understand a perturbation expression of the
fine-grained entropy and to investigate the behavior of coarse-grained
entropy.

\subsection{Fine-grained Entropy}

We use the standard expression for Maxwell-Boltzmann entropy,
\begin{equation}\label{fgs}
s=- \int \int f \ln{f} \,
\dif \chi \dif \varpi.
\end{equation} 
As mentioned previously, we have also utilized a Lynden-Bell entropy.
As our situations are not degenerate, there is essentially no
difference between values derived from the two approaches, and we will
simply refer to entropy in this discussion.  Using the perturbation
expansion in Equation~\ref{pertf} allows us to write the fine-grained
entropy to second-order as,
\begin{eqnarray}\label{pfgs0}
\lefteqn{s = s_0 + \epsilon s_1 + \epsilon^2 s_2 =} & & \\
& & -\iint f_0 \ln{f_0} \, \dif \chi \dif \varpi -\epsilon \iint
f_1(1+\ln{f_0}) \, \dif \chi \dif \varpi - \nonumber \\
& & \epsilon^2 \iint \left[ f_2(1+\ln{f_0}) +
\frac{f_1^2}{2f_0} \right] \, \dif \chi \dif \varpi.
\end{eqnarray}
Due to the Boltzmann form of equilibrium, we have that
\begin{equation}
\ln{f_0}=-\ln{(2\sqrt{\pi})} - e_0 = -\ln{(2\sqrt{\pi})} -
\varpi^2 - 2\ln{(2 \cosh{\chi})},
\end{equation}
where $e$ is the dimensionless energy per unit mass.  In order to
complete the integrations, we need to take advantage of the fact that,
\begin{eqnarray*}
\lefteqn{\int_{-\infty}^{\infty} P_n(\tanh \chi) \ln{(\cosh \chi)} 
\sech^2 \, \dif \chi =} & & \\
& & -\frac{1}{2}\int_{-1}^1 P_n(u) \ln{(1-u^2)} \, \dif u = \left\{ 
\begin{array}{l} 2(1-\ln{2}), \, \mbox{for $n=0$} \\
\frac{2}{n(n+1)}, \, \mbox{for $n\ge 2$, even.}
\end{array} \right.
\end{eqnarray*}
The zeroth-, first-, and second-order fine-grained entropy expressions
can be written as,
\begin{eqnarray}\label{pfgs}
s_0 & = & \ln{(2\sqrt{\pi})} + \frac{5}{2}, \nonumber \\
s_1 & = & 4\sqrt{\pi} \left[ c_{2,0} + \sum_{n \ge 2 \atop {\rm
even}} \frac{c_{0,n}}{n(n+1)} \right], \\
s_2 & = & 4\sqrt{\pi} \left[ d_{2,0} + \sum_{n \ge 2 \atop {\rm
even}} \frac{d_{0,n}}{n(n+1)} \right] - 2\pi \sum_{m,n\atop m\ne n=0} 
\frac{2^m m!}{2n+1} c_{m,n}^2. \nonumber
\end{eqnarray}
The first expression in Equation~\ref{pfgs} is trivially
time-independent, and $\dot{s}_1=0$ because that term is the same as
the first-order energy (ignoring the even-odd coefficients), which is
time-independent.  We note that final term in the $s_2$
expression shows that any first-order perturbation (all $d_{m,n}=0$)
results in a reduction in entropy.  As long as the perturbation
imparts no first-order energy ($E_1=s_1=0$), the separable equilibrium
is also the maximum entropy state.

To show that $\dot{s}_2=0$, we use Equation~\ref{de1} to write,
\begin{eqnarray}\label{ds20}
\dot{s}_2 & = & -8\pi \sum_{n\ge 1} \frac{c_{0,n}}{2n+1} \left[
\frac{c_{1,n+1}}{2n+3}-\frac{c_{1,n-1}}{2n-1} \right] - \nonumber \\
& & 4\pi \sum_{m,n\atop m\ne n=0} \frac{2^m m!}{2n+1} 
c_{m,n}\dot{c}_{m,n}.
\end{eqnarray}
We focus our attention on the second right-hand-side term.  Using the
first-order coefficient dynamics equations, we see that these
$c_{m,n}\dot{c}_{m,n}$ terms behave just like coefficients in the
test-particle case, except for $m=1$.  The rightmost term in
Equation~\ref{ds20} can be re-cast as,
\begin{eqnarray}
\lefteqn{\sum_{m,n\atop m\ne n=0} \frac{2^m m!}{2n+1}
c_{m,n}\dot{c}_{m,n}=
\sum_{m,n\atop m\ne n=0} \frac{2^m m!}{2n+1} c_{m,n}^{\rm test}
\dot{c}_{m,n}^{\rm test} -} & & \nonumber \\
& & \sum_{n\ge 0} \frac{2c_{1,n}}{2n+1} \left[
\frac{c_{0,n-1}}{2n-1} -\frac{c_{0,n+1}}{2n+3} \right].
\end{eqnarray}
We show that the test-particle term is zero in Appendix~\ref{testdem}.  
In order to demonstrate the time-independence of $s_2$, we need to
re-define the index variables in the last term of this expression.
Taking $k=n-1$, we re-write
\begin{equation}
\sum_{n\ge 0} \frac{c_{1,n}c_{0,n-1}}{(2n+1)(2n-1)} = \sum_{k \ge
1} \frac{c_{1,k+1}c_{0,k}}{(2k+3)(2k+1)},
\end{equation}
where the $k=-1$ and $k=0$ contributions disappear since $c_{0,-1}$
and $c_{0,0}$ are both zero.  Similarly, taking $k=n+1$, we re-write
\begin{equation}
\sum_{n\ge 0} \frac{c_{1,n}c_{0,n+1}}{(2n+1)(2n+3)} = \sum_{k \ge
1} \frac{c_{1,k-1}c_{0,k}}{(2k-1)(2k+1)}.
\end{equation}
With these expressions, Equation~\ref{ds20} becomes
\begin{eqnarray}
\dot{s}_2 & = & -8\pi \sum_{n\ge 1} \frac{c_{0,n}}{2n+1} \left[
\frac{c_{1,n+1}}{2n+3}-\frac{c_{1,n-1}}{2n-1} \right] + \nonumber \\
& & 8\pi \sum_{k\ge 1} \frac{c_{0,k}}{2k+1} \left[
\frac{c_{1,k+1}}{2k+3} - \frac{c_{1,k-1}}{2k-1} \right] = 0.
\end{eqnarray}

The time-independence of our fine-grained entropy gives us confidence
that we have meaningful expressions, and numerical simulations of
coefficient evolutions (see \S~\ref{coeffsims}) verify that these are
conserved quantities.  Assuming that Equation~\ref{pfgs} correctly
describes the fine-grained entropy, we make several observations.
One, first-order entropy is identical with first-order energy.  Two,
second-order, fine-grained entropy is held constant by an interplay
between terms associated with second-order energy (those in square
brackets) and $c_{m,n}^2$ terms arising from the self-interaction of
the perturbation.  Three, if a given perturbation does not populate
the terms in brackets in Equation~\ref{pfgs}, then the negative sign
on the $c_{m,n}^2$ term guarantees that the perturbed system entropy
must be lower than the equilibrium $s_0$ value.  This sets up the
possibility that coarse-grained entropy could increase back to the
equilibrium value, given appropriate initial conditions.  

\subsection{Coarse-grained Entropy}

\subsubsection{Perturbation Analysis}

The fine-grained entropy is time-independent because the fine-grained
distribution function obeys the collisionless Boltzmann equation.  In
the absence of a relaxation mechanism (like collisions), there can be
no entropy creation or destruction.  This changes when one
investigates a coarse-grained distribution function.  In contrast to
quantum systems where Planck's constant provides a natural phase-space
benchmark, coarse-graining is an ill-defined procedure for classical
systems like the ones we are discussing.  We take advantage of this
freedom by using two coarse-graining definitions.

We define our first, and more standard, coarse-graining procedure as
taking an average of the fine-grained distribution function over some
range in position and velocity,
\begin{equation}\label{cg1}
F_{xv}(\chi,\varpi)= \frac{1}{\Delta \chi \Delta \varpi}
\int_{\chi_1}^{\chi_2} \int_{\varpi_1}^{\varpi_2}
f(\chi^{\prime},\varpi^{\prime}) \, \dif \chi^{\prime} \dif
\varpi^{\prime},
\end{equation}
where $\chi_1=\chi-\Delta \chi/2$, $\chi_2=\chi+\Delta \chi/2$,
$\varpi_1=\varpi - \Delta \varpi/2$, and $\varpi_2=\varpi + \Delta
\varpi/2$.  A more straightforward coarse-graining can be
obtained directly in $(m,n)$-space by simply truncating the
fine-grained distribution function expansion,
\begin{equation}\label{cg2}
F_{mn}(\chi,\varpi) = \sum_{m=0}^M \sum_{n=0}^N c_{m,n} H_m(\varpi)
P_n(\tanh \chi) \sech^2 \chi e^{-\varpi^2},
\end{equation}
where $M$ and $N$ are integers greater than 2.  This coarse-graining
simply does not allow small-scale position and velocity features
(represented by large $m$ and $n$ terms) to be represented in the
distribution function.

In general, these choices allow us to write the coarse-grained
distribution function in terms of the fine-grained function and what
we will call a relaxation function $\gamma$,
\begin{equation}\label{cgf}
F_i = f + \gamma_i,
\end{equation}
where the subscript distinguishes between the different
coarse-graining schemes.  Using Equation~\ref{fexpand}, we
integrate Equation~\ref{cg1} to produce,
\begin{eqnarray}\label{xvcg1}
\lefteqn{F_{xv} = \frac{1}{\Delta \chi \Delta \varpi} \sum_{m,n}
\frac{c_{m,n}}{2n+1} \times } & & \nonumber \\
& & \left\{ H_{m-1}(\varpi_1)e^{-\varpi_1^2} -
H_{m-1}(\varpi_2)e^{-\varpi_2^2} \right\} \times \nonumber \\
& & \left\{ \left[ P_{n+1}(\tanh \chi_2) -
P_{n+1}(\tanh \chi_1)\right]- \right. \nonumber \\
& & \left. \left[ P_{n-1}(\tanh \chi_2) - P_{n-1}(\tanh \chi_1)
\right] \right\}.
\end{eqnarray}
The various Hermite and Legendre polynomials in this expression can be
expanded about $\varpi$ and $\chi$, respectively.  Taking advantage of
the fact that,
\begin{equation}
\frac{\dif^p}{\dif \varpi^p} [ H_{m-1}(\varpi) e^{-\varpi^2}] = (-1)^p
H_{m+p}(\varpi) e^{-\varpi^2},
\end{equation}
lets us re-write the difference in Hermite terms in
Equation~\ref{xvcg1} as,
\begin{eqnarray}
\lefteqn{H_{m-1}(\varpi_1)e^{-\varpi_1^2} -
H_{m-1}(\varpi_2)e^{-\varpi_2^2} \approx} & & \\
& & H_m(\varpi) e^{-\varpi^2} \Delta \varpi + H_{m+2}(\varpi)
e^{-\varpi^2} \left( \frac{\Delta \varpi^3}{24}\right) +
\mathcal{O}(\Delta \varpi^5). \nonumber
\end{eqnarray}
Expanding the Legendre polynomials allows us to write,
\begin{eqnarray}
\lefteqn{F_{xv} \approx f + \sum_{m,n}
\frac{c_{m,n}}{2n+1} H_m(\varpi) \times} & & \nonumber \\
& & \frac{\dif^3}{\dif \chi^3} \left[
P_{n+1}(\tanh \chi) - P_{n-1}(\tanh \chi) \right] e^{-\varpi^2}
\left( \frac{\Delta \chi^2}{24} \right) + \nonumber \\
& & \sum_{m,n} c_{m,n} H_{m+2}(\varpi) P_n(\tanh \chi) \sech^2 \chi
e^{-\varpi^2} \left( \frac{\Delta \varpi^2}{24} \right),
\end{eqnarray}
accurate to second-order in the coarse-graining sizes.  Taking $\Delta
\varpi=\Delta \chi=\Delta$ and performing the $\chi$-differentiation
results in,
\begin{eqnarray}\label{xvcg2}
\lefteqn{\gamma_{xv} = \frac{\Delta^2}{24} \sum_{m,n} c_{m,n}
\left\{ A_n H_m(\varpi) P_{n+2}(\tanh \chi) + \right.} & & \nonumber
\\
& & [B_n + C_n P_2(\tanh \chi)] H_m(\varpi) P_n(\tanh \chi) +
\nonumber \\
& & \left. H_{m+2}(\varpi) P_n(\tanh \chi) \right\} \sech^2 \chi 
e^{-\varpi^2},
\end{eqnarray}
where 
\begin{eqnarray*}
A_n & = & [4(n+1)(n+2)]/(2n+3), \\
B_n & = & [-2n(n+1)(2n+1)]/(3(2n+3)), \; \mbox{and} \\
C_n & = & [2(n-1)(n-2)]/3.
\end{eqnarray*}
We note that integrating this expression for the coarse-grain
relaxation function over all of phase space results in zero, leaving
the total mass of the system unchanged.

Coarse-graining in $(m,n)$-space results in a relaxation
function with the form,
\begin{equation}\label{mncg1}
\gamma_{mn} = \gamma_A + \gamma_B + \gamma_C,
\end{equation}
where
\setlength{\arraycolsep}{1mm}
\begin{eqnarray*}
\gamma_A & = & -\sum_{m= \atop M+1}^{\infty} 
\sum_{n= \atop N+1}^{\infty} c_{m,n}
H_m(\varpi) P_n(\tanh \chi) \sech^2 \chi e^{-\varpi^2}, \\
\gamma_B & = & -\sum_{m=0}^{M} \sum_{n=N}^{\infty} c_{m,n}
H_m(\varpi) P_n(\tanh \chi) \sech^2 \chi e^{-\varpi^2}, \; 
\mbox{and} \\
\gamma_C & = & -\sum_{m= \atop M+1}^{\infty} \sum_{n=0}^{N} c_{m,n}
H_m(\varpi) P_n(\tanh \chi) \sech^2 \chi e^{-\varpi^2}.
\end{eqnarray*}
These terms represent the behavior of the distribution function on
scales smaller than the coarse-graining size, which is determined by
the choice of $M$ and $N$.  We do not define a specific relationship
between $M$ and $N$ and $\Delta \chi$ and $\Delta \varpi$.   However,
since $M$ and $N$ represent the numbers of roots of polynomials,
larger values roughly correspond to smaller $\Delta \chi$ and $\Delta
\varpi$ values.

Regardless of the specific approach to coarse-graining, one can define
a dimensionless coarse-grained entropy along the same lines as
Equation~\ref{fgs},
\begin{equation}\label{cgs}
S=- \int \int F \ln{F} \, \dif \chi \dif \varpi.
\end{equation}
Starting from Equation~\ref{cgf}, we assume that $\gamma \ll f$ at any
$(\chi,\varpi)$.  For $\gamma_{xv}$, this amounts to assuming that the
coarse-graining kernel size $\Delta^2$ is small.  It is not as
obvious that this condition is satisfied by $\gamma_{mn}$, however the
oscillatory nature of the coefficient values makes it reasonable
to expect a linear combination of such values to remain relatively
small.  We will show that this assumption is justified in a later
section.  Upon expansion of the logarithm in
Equation~\ref{cgs} we find,
\begin{equation}
F \ln{F} \approx f \ln{f} + \gamma(1+\ln{f}) +
\frac{\gamma^2}{2f},
\end{equation}
which is accurate to second-order in $\gamma$.  If we then use the
perturbation expansion of $f=f_0 + \epsilon f_1 + \epsilon^2
f_2$ and a corresponding expansion of $\gamma = \epsilon \gamma_1 +
\epsilon^2 \gamma_2$ (no zeroth-order correction is needed), we
then write,
\begin{eqnarray}\label{cgflnf}
\lefteqn{F \ln{F} \approx f_0 \ln{f_0} +} & & \nonumber \\
& & \epsilon\left[ f_1(1+\ln{f_0)} + \gamma_1(1+\ln{f_0})
\right] + \nonumber \\
& & \epsilon^2 \left[ f_2(1+\ln{f_0}) + \gamma_2(1+\ln{f_0}) +
\frac{(f_1 + \gamma_1)^2}{2f_0} \right],
\end{eqnarray}
which is accurate to second-order in the perturbation strength.
Several terms are familiar from Equation~\ref{pfgs0}, so we will focus
on the additions due to coarse-graining.  Note that both the first-
and second-order $\gamma$ terms are composed of the three pieces in
Equation~\ref{mncg1}.  Those expressions following
Equation~\ref{mncg1} combine to form $\gamma_1$, while substituting
$d_{m,n}$ for $c_{m,n}$ in those formulae lead to $\gamma_2$.

Both coarse-graining prescriptions produce zero-mass first-order
perturbations,
\begin{equation}
\int \int \gamma_1 \, \dif \chi \dif \varpi = 0.
\end{equation}
The $(x,v)$ coarse-graining correction to the first-order entropy
is,
\begin{eqnarray}\label{focgs1}
\lefteqn{\iint \gamma_{xv,1} (1+\ln{f_0}) \, \dif \chi \dif \varpi =
-\sqrt{\pi}\frac{\Delta^2}{6} \sum_{n\ge 2 \atop {\rm even}} c_{0,n}
\times } & & \nonumber \\
& & \left\{ \frac{A_n}{(n+2)(n+3)} + \frac{B_n}{n(n+1)} + \right. \\ 
& & 4C_n \left[ \frac{3(n+1)^2(n+2)}{2(2n+4)(2n+3)(2n+2)(2n+1)(n+3)} +
\right. \nonumber \\
& & \left. \left. \frac{n(n+1)}{2n(2n+2)(2n+1)(2n-1)} \right] +
\frac{4n(n-1)^2}{3(2n-1)(2n-2)(2n-3)} \right\}. \nonumber
\end{eqnarray}
For contrast, the $(m,n)$ coarse-graining correction to the
first-order entropy is,
\begin{equation}\label{focgs2}
\iint \gamma_{mn,1} (1+\ln{f_0}) \, \dif \chi \dif \varpi =
4\sqrt{\pi} \sum_{n \ge N+1 \atop {\rm even}}
\frac{c_{0,n}}{n(n+1)}.
\end{equation}
Note that this has the same form as the first-order fine-grained
entropy expression (Equation~\ref{pfgs}), with different summation
limits.  We will show that this trend continues with the second-order
expressions, making the similarity between fine-grained and
coarse-grained entropy a major advantage of adopting the $(m,n)$
procedure.

For the second-order entropy, we will confine ourselves to a
discussion of the $(m,n)$ coarse-graining prescription.  Details of
the $(x,v)$ route are given in Appendix~\ref{xvcgs2}.  As with the
first-order coarse-grained perturbation, the second-order perturbation
is massless.  There are three terms that need to be explored.  The first
one is exactly analogous to the first-order term,
\begin{equation}
\iint \gamma_{mn,2} (1+\ln{f_0}) \, \dif \chi \dif \varpi =
4\sqrt{\pi} \sum_{n \ge N+1 \atop {\rm even}}
\frac{d_{0,n}}{n(n+1)},
\end{equation}
where the second-order perturbation coefficients $d_{0,n}$ have taken
the place of the first-order coefficients.  The next involves the
first-order coarse-grained perturbation squared and is,
\begin{eqnarray}
\iint \frac{\gamma_{mn,1}^2}{f_0} \, \dif \chi \dif \varpi & = & 4\pi
\sum_{m\ge M+1} \sum_{n\ge N+1} c_{m,n}^2 \frac{2^m m!}{2n+1} +
\nonumber \\
& & 4\pi \sum_{m=0}^M \sum_{n\ge N+1} c_{m,n}^2 \frac{2^m m!}{2n+1}
+ \nonumber \\
& & 4\pi \sum_{m \ge M+1} \sum_{n=0}^N c_{m,n}^2 \frac{2^m m!}{2n+1}.
\end{eqnarray}
Finally, the term involving the product of the first-order fine- and
coarse-grained functions is,
\begin{eqnarray}
\iint \frac{f_1 \gamma_{mn,1}}{f_0} \, \dif \chi \dif \varpi & = & 
-4\pi \sum_{m\ge M+1} \sum_{n\ge N+1} c_{m,n}^2 \frac{2^m m!}{2n+1} -
\nonumber \\
& & 4\pi \sum_{m=0}^M \sum_{n\ge N+1} c_{m,n}^2 \frac{2^m m!}{2n+1} -
\nonumber \\
& & 4\pi \sum_{m\ge M+1} \sum_{n=0}^N c_{m,n}^2 \frac{2^m m!}{2n+1}.
\end{eqnarray}
Taken together, the coarse-grain second-order entropy is,
\begin{eqnarray}
S_{mn,2} & = & 
4\sqrt{\pi} \left( d_{2,0} + \sum_{n=2 \atop {\rm even}}^N
\frac{d_{0,n}}{n(n+1)} \right) - \nonumber \\
 & & 2\pi \sum_{m=0}^M \sum_{n=0 \atop m\ne n=0}^N c_{m,n}^2 
\frac{2^m m!}{2n+1}.
\end{eqnarray}
As with the first-order entropy, the $(m,n)$ coarse-graining results
in an entropy expression that mirrors the fine-grained expression,
apart from the limits of the summations.

The coarse-grained entropy up to second-order for a perturbed system
can be now written as,
\begin{eqnarray}\label{cgs0}
S_{mn} & = & \ln{(2\sqrt{\pi})} + \frac{5}{2} + \nonumber \\
& & 4\sqrt{\pi} \epsilon \left( c_{2,0} + 
\sum_{n 2 \atop {\rm even}}^N \frac{c_{0,n}}{n(n+1)} \right) +
\nonumber \\
& & \epsilon^2 \left[ 4\sqrt{\pi} \left( d_{2,0} + 
\sum_{n=2 \atop {\rm even}}^N \frac{d_{0,n}}{n(n+1)} \right) -
\right. \nonumber \\
& & \left. 2\pi \sum_{m=0}^M \sum_{n=0 \atop m\ne n=0}^N c_{m,n}^2 
\frac{2^m m!}{2n+1} \right].
\end{eqnarray}
The time rate of change (denoted by a dot where simple) of this
coarse-grained entropy is simply,
\begin{eqnarray}\label{dcgs1}
\dot{S}_{mn} & = & 4\sqrt{\pi} \epsilon \left( \dot{c}_{2,0} + 
\sum_{n \ge 2 \atop {\rm even}}^N \frac{\dot{c}_{0,n}}{n(n+1)}
\right)+ \nonumber \\
& & \epsilon^2 \left[ 4\sqrt{\pi} \left( \dot{d}_{2,0} + 
\sum_{n=2 \atop {\rm even}}^N \frac{\dot{d}_{0,n}}{n(n+1)} \right) -
\right. \nonumber \\
& & \left. 2\pi \frac{\partial}{\partial \tau} 
\sum_{m=0}^M \sum_{n=0 \atop m\ne n=0}^N c_{m,n}^2 \frac{2^m m!}{2n+1} 
\right].
\end{eqnarray}

Equations~\ref{cgs0} and \ref{dcgs1} provide us with a straightforward
conceptual picture of the behavior of coarse-grained entropy.  Any
initial first-order perturbation populates a set of $c_{m,n}$ values
and defines an initial entropy (all initial $d_{m,n}=0$).  Coefficient
dynamics demand that most first-order coefficient values diminish, or
vanish, in the wake of a ``wave'' that propagates to ever larger $m$
and $n$ values \citep{br14}.  This occurs for both test-particle and
self-gravitating systems.  This behavior is phase mixing seen in
coefficient space.  Initial, large-scale $(x,v)$ perturbations [small
$(m,n)$] become reversibly transformed to small-scale $(x,v)$
perturbations [large $(m,n)$].

In situations where only coefficients with $m \le M$ and $n \le N$ are
initially populated, the radiation-like behavior of the coefficients
demands that the last term in Equation~\ref{dcgs1} provide a positive
contribution to the entropy change.  This happens due to the
conservation of fine-grained entropy, with coefficient values inside
the $M$ by $N$ coarse-graining region decreasing as coefficients with
$m > M$ and $n > N$ must become non-zero.

If, on the other hand, coefficients with $m > M$ and $n >
N$ are initially populated, then the coefficient dynamics will cause a
decrease in the coarse-grained entropy as coefficients with $m \le M$
and $n \le N$ become non-zero as part of the ``wave'' that radiates
inwards from larger indices.  However, in this situation, the reflecting
nature of the $m=0$ and $n=0$ boundaries eventually causes the
coarse-grained entropy to increase as any populated low $m,n$
coefficients will then behave as in the situation above.

\subsubsection{Coefficient Dynamics Simulations}\label{coeffsims}

It might be expected, then, that first-order coarse-grained entropy
should increase.  In order to test this expectation, we have
numerically solved the first- and second-order coefficient dynamics
equations (see discussion of Equation~\ref{cbelp} in
Section~\ref{pertsec}).  A simple midpoint method integration with
fixed time step is used to solve for coefficient values on an ($m,n$)
grid that has reflecting boundaries.  A more detailed discussion of
the performance of these types of integrations and boundary conditions
is presented in \citet{br14}.  The second-order calculation is far
more expensive than the first-order calculation, and as a result we
have limited our calculations to grids with $m_{\rm max}=n_{\rm
max}=49$.  With a parallel calculation of $R_{m,n}$ and a time step of
$\approx 2\times 10^{-3}$ on a standard multi-core processor machine,
coefficient evolutions for 10 time units take approximately 20 hours.
The grid size also limits the duration of the evolution, as
reflections affect low ($m,n$) coefficient evolutions after
sufficiently long times.

\begin{figure}
\scalebox{0.5}{
\includegraphics{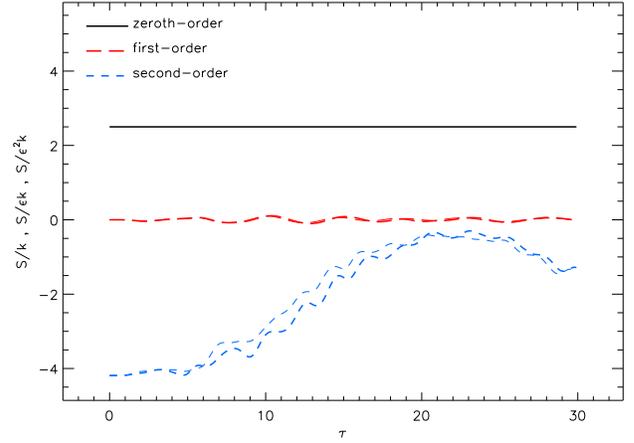}}
\caption{Evolution of the $(m,n)$ coarse-grained entropy contributions
for an initial $c_{1,1}$ perturbation.  To magnify the behaviors, each
set of values has been divided by the corresponding perturbation
strength.  Thin lines illustrate test-particle values, while thick
lines represent self-gravitating values.  First-order entropy
behaviors are marked with long dashed lines.  Second-order entropy
evolutions are marked with short dashed lines.  The first-order curves
remain nearly constant and show no differences between test-particle
and self-gravitating systems.  The second-order curves show noticeable
increases, with the late-time decreases likely stemming from
numerical issues.  Differences between test-particle and self-gravitating
systems are noticeable at this order, but the overall behaviors are
rather similar.
\label{cgs11}}
\end{figure}

\begin{figure}
\scalebox{0.5}{
\includegraphics{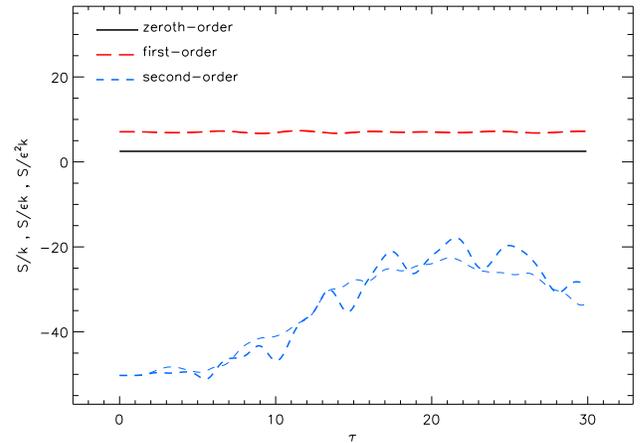}}
\caption{Evolution of the $(m,n)$ coarse-grained entropy contributions
for an initial $c_{2,0}$ perturbation.  As in Figure~\ref{cgs11}, the
values for the various orders have been normalized by the appropriate
perturbation strength.  Again, the behaviors of the test-particle
(thin) and self-gravitating (thick) curves are very similar at every
order.
\label{cgs20}}
\end{figure}

These numerical evaluations indicate that coefficient values and their
time derivatives tend to have oscillating values that result in
first-order coarse-grained entropies that oscillate about their
fine-grained value.  Only when one looks at second-order
coarse-grained entropy can an overall increase be seen.  This is
evident in Figure~\ref{cgs11}, which shows first- and second-order
coarse-grained entropy evolutions that result from an initial $c_{1,1}$
perturbation.  We note that the modest decreases in average
entropy values -- \ie, the non-oscillatory changes that
occur after $\tau \approx 25$ -- in Figures~\ref{cgs11}, \ref{cgs20}, and
\ref{cgs02} likely indicate the impact of boundary-induced
reflections in the numerical scheme.

Test-particle systems can only experience phase mixing, but perturbed
self-gravitating systems have the possibility of experiencing violent
relaxation in addition to phase mixing.  In an attempt to disentangle
the impact of each process on the entropy evolution, we compare
coarse-grained entropy behaviors in test-particle and self-gravitating
systems subject to identical perturbations.  As shown in
Figures~\ref{cgs11} and \ref{cgs20}, there do not appear to be
significant or systematic differences between the behaviors in
test-particle and self-gravitating situations.  At least for the
modest perturbation strengths investigated here, violent relaxation
has a much smaller impact on entropy creation compared to phase
mixing.

The behavior of the entropy is in line with the Tremaine \etal\
prediction that the coarse-grained entropy should not decrease from
its initial value.  The multiple, competing, terms present in
Equation~\ref{dcgs1} are an expression of why a stronger statement
cannot be made, like the guarantee of monotonic increase for
collisional systems.  The time-derivative of first-order entropy can
be positive or negative, and while second-order entropy shows an
overall increase during relaxation due to phase mixing, it is not a
monotonic increase.

\subsubsection{Non-linear Perturbation Simulations}

Our perturbation analysis has led us to speculate that violent
relaxation is not a significant source of entropy production.  We have
investigated whether or not this is simply an artifact of our
limitations on perturbation strength.  Using $N$-body simulations
\citep[for numerical method details see] []{br14}, we have also
explored entropy evolutions in self-gravitating and test-particle
systems that are so far from equilibrium that our perturbation
expressions are not appropriate.  These initial conditions are
``waterbags'' \citep{j11} with different amounts of kinetic energy.
For these $N$-body simulations, entropy is calculated  using a
particle counting scheme,
\begin{equation}
S_{\rm NB} = - \sum_i n_i \ln{n_i},
\end{equation}
where $n$ is particle number and $i$ enumerates different areas of
phase space (all of size $\Delta \chi \Delta \varpi$).  Unlike in
quantum situations where $\Delta \chi$ and $\Delta \varpi$ can be
related to Planck's constant, we have simply used trial and error to
set sizes of the phase-space boxes.  After investigating a wide range,
we have found that values near the adopted $\Delta \chi = \Delta
\varpi = 2\times 10^{-2}$ produce entropy values that show the most
obvious changes during evolution.  Smaller values result in almost no
particles falling into the boxes, while larger values produce boxes so
large that variation is basically absent.  In either case, resulting
entropy changes are small.

\begin{figure}
\scalebox{0.5}{
\includegraphics{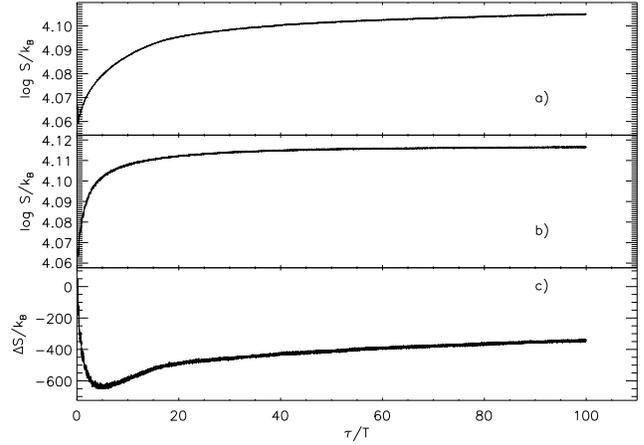}}
\caption{Comparisons between ensemble average entropy behaviors in
self-gravitating (panel a) and test-particle (panel b) $N$-body
simulations starting from identical initial conditions.  Particles are
placed spatially according to a uniform distribution of random values.
There is no initial kinetic energy, making this an extremely
non-equilibrium distribution.  That the test-particle entropy shows a
larger change than in the self-gravitating case can be seen in panel
c, where the self-gravitating entropy value minus the test-particle
entropy value is shown as a function of time.
\label{bar0}}
\end{figure}

\begin{figure}
\scalebox{0.5}{
\includegraphics{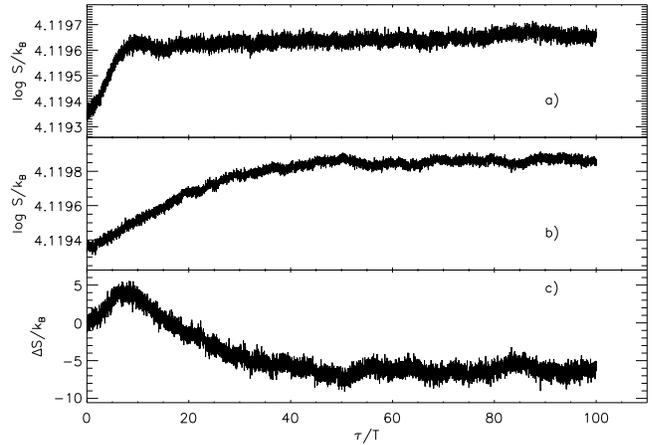}}
\caption{Analogous to Figure~\ref{bar0}, but systems are given half of
the virial equilibrium kinetic energy.  This rectangular phase-space
distribution is much closer to approximating equilibrium compared to
the zero kinetic energy cases.  Note that the overall changes to
self-gravitating (panel a) and test-particle (panel b) entropies are
greatly diminished compared to those in Figure~\ref{bar0}.
Again, panel c shows that the test-particle entropy shows a
larger change than in the self-gravitating case.
\label{bar1}}
\end{figure}

Somewhat surprisingly, test-particle systems experience larger entropy
changes during relaxation, as shown in Figure~\ref{bar0}.  We note
that the maximum difference between test-particle and self-gravitating
entropies is approximately 5\% of either entropy value, again
indicating that violent relaxation does not have a strong impact on
entropy behavior.  In agreement with the second-order results,
differences between self-gravitating and test-particle systems
disappear as the non-linearity of initial conditions decreases.
Figure~\ref{bar1} shows how much smaller the entropy differences are
when the system has half of the kinetic energy required for virial
equilibrium and consequently undergoes a much milder relaxation.

\subsubsection{Quantifying Incomplete Relaxation}

Any perturbation involving only first-order coefficients leads to
predictable time-independent modes \citep{rb19a}.  These
time-independent sets of coefficient values lead to upper-limits on
changes that coarse-grained entropy can experience.  For example, any
odd-$m$, odd-$n$ perturbation results in a time-independent mode where
the only non-zero coefficient values exist at $m=n=\infty$.  The loss
of all of the structure information encoded in the coefficients
results in a maximal gain in coarse-grained entropy.  However, for
perturbations that leave residual time-independent coefficients, the
coarse-grained entropy value increases or decreases with the values of
$M$ and $N$.  Smaller coarse-graining boxes lead to smaller increases
in entropy.

We have numerically determined time-independent modes, or sets of
coefficients, and used them to calculate the maximum coarse-grained
entropy change resulting from perturbations by individual $(m,n)$
coefficients.  Behavior of odd-$m$, odd-$n$ perturbations has been
described above, so we focus on even-$m$, even-$n$ perturbations.  As
the perturbing $m$ and $n$ values increase, the entropy change
approaches the maximal value associated with complete relaxation.
More physically, these curves show that perturbations with larger
position and velocity scales (smaller $n$ and $m$ values) undergo
substantially more incomplete relaxation.

\begin{figure}
\scalebox{0.5}{
\includegraphics{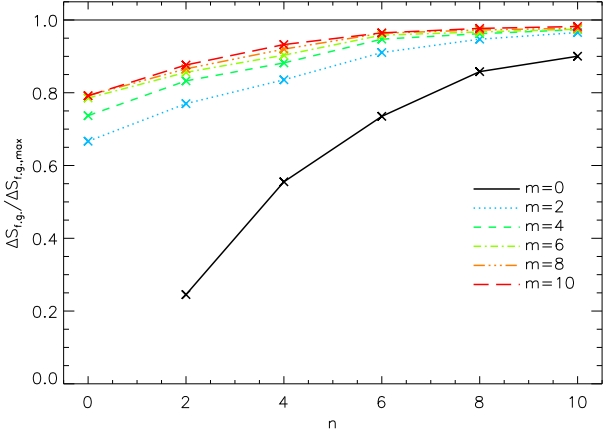}}
\caption{Differences in second-order fine-grained entropy between
initial conditions and time-independent modes for several
even-$m$, even-$n$ perturbations.  As either $m$ or $n$ increase, the
difference increases.  A $(0,2)$ perturbation holds onto the maximum
amount of its initial entropy during its evolution; it relaxes the
smallest amount.
\label{eeds2}}
\end{figure}

\begin{figure}
\scalebox{0.5}{
\includegraphics{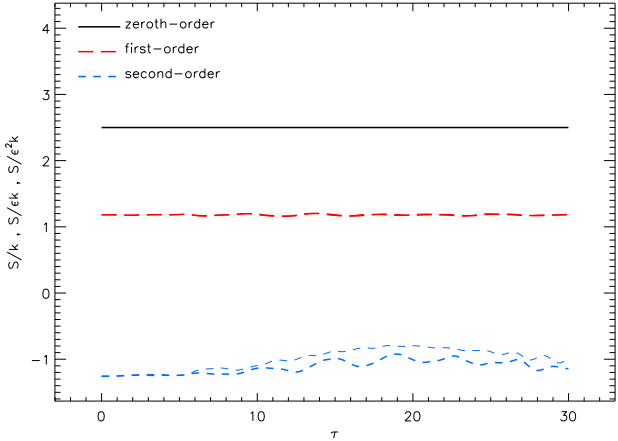}}
\caption{Evolution of the $(m,n)$ coarse-grained entropy contributions
for an initial $c_{0,2}$ perturbation, with the same curve
descriptions as in Figures~\ref{cgs11} and \ref{cgs20}.  The entropy
change present in this plot is noticeably smaller than those in
Figures~\ref{cgs11} and \ref{cgs20}, indicating that the perturbation
is more nearly a time-independent mode which can relax little.
\label{cgs02}}
\end{figure}

Qualitatively, the relationship between the entropy differences for
$(m=2,n=0)$ and $(m=0,n=2)$ seen in Figure~\ref{eeds2} agrees with the
coarse-grained entropy changes seen in Figures~\ref{cgs20} and
\ref{cgs02}.  A $c_{0,2}$ perturbation is a significant contribution
to a time-independent mode.  As a result, there is relatively little
relaxation that is possible.

As each time-independent mode corresponds to a unique amount of
entropy, we think of the collection of these mode strengths as
quantifying the amount of relaxation possible.  If a given
perturbation does not contain any time-independent modes, then a
complete relaxation is possible.  It is important to note that these
kinds of perturbations also contain zero energy, so the system will
return to its underlying separable equilibrium.  Perturbations
that are energetic must contain time-independent modes and preclude
complete relaxation.  As in Figures~\ref{cgs20} and \ref{cgs02}, the
particular modes determine how incomplete the relaxation will be.

\subsection{Thermodynamic Entropy}

We have also taken a more standard thermodynamic approach to
calculating the entropy in these models \citep{dgm85}.  To begin, we
list relevant moments of the collisionless Boltzmann equation:
\begin{eqnarray*}
\mbox{zeroth moment,} \quad \frac{\partial}{\partial \tau}[ \Lambda ]
& = & -\frac{\dif}{\dif \chi} \left[ \Lambda \avgv \right] \\
\mbox{first moment,} \quad \frac{\partial}{\partial \tau} [ \Lambda 
\avgv ]
& = & -\frac{\dif}{\dif \chi} \left[ \Lambda \avgvv \right] - 
\Lambda \frac{\dif \phi}{\dif \chi} \\
\mbox{second moment,} \quad \frac{\partial}{\partial \tau}
\left[ \frac{\Lambda}{2} \avgvv  \right] & = & -\frac{\dif}{\dif \chi}
\left[ \frac{\Lambda}{2} \avgvvv \right] - \Lambda \avgv 
\frac{\dif \phi}{\dif \chi}.
\end{eqnarray*}

Our goal is to use the first law of thermodynamics,
\begin{equation}
\dif u = T \dif s - p \; \dif L,
\end{equation}
where $u$ is internal energy per unit mass, $T$ is temperature, $p$ is
pressure, and $L$ is the extent of the system (the one-dimensional
analogue to volume $V$), to understand how entropy changes.  Keeping
with typical definitions,
\begin{equation}
p \equiv \Lambda(\avgvv - \avgv^2),
\end{equation}
and
\begin{equation}
T \equiv \avgvv.
\end{equation}
We use the angle brackets to denote average quantities that are
calculated as
\begin{equation}
\langle A \rangle \equiv \frac{1}{\Lambda} \int_{-\infty}^{\infty} A f
\; \dif \varpi.
\end{equation}
With these definitions, we find that
\begin{eqnarray}\label{flt0}
\lefteqn{\Lambda \frac{\dif}{\dif \tau} \left[ \frac{\avgvv - \avgv^2}{2} 
\right] = -p\frac{\dif \avgv}{\dif \chi} -} \nonumber \\
 & & \frac{\dif}{\dif \chi}
\left[ \frac{\Lambda}{2} ( \avgvvv - 3 \avgvv \avgv + 2 \avgv^3 
\right].
\end{eqnarray}
Using the continuity equation (the zeroth moment from above), we can
re-write the first term on the right-hand side of this expression as
\begin{equation}
\frac{\dif \avgv}{\dif \chi} = \Lambda \frac{\dif}{\dif \tau} \left(
\frac{1}{\Lambda} \right) = \Lambda \frac{\dif L}{\dif \tau}.
\end{equation}
Identifying the internal energy per unit mass as,
\begin{equation}
u \equiv \frac{\avgvv - \avgv^2}{2},
\end{equation} 
we can now re-cast Equation~\ref{flt0} as
\begin{equation}\label{flt1}
\frac{\dif u}{\dif \tau} =
-p\frac{\dif L}{\dif \tau} - \frac{1}{\Lambda} \frac{\dif}{\dif \chi}
\left[ \frac{\Lambda}{2} ( \avgvvv - 3 \avgvv \avgv + 2 \avgv^3 
\right].
\end{equation}
This last term must be equal to $T \dif s/\dif \tau$, according to the
first law of thermodynamics.

\begin{figure}
\scalebox{0.5}{
\includegraphics{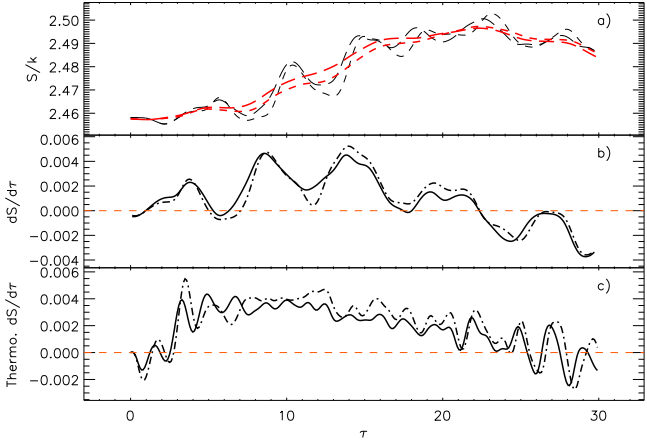}}
\caption{Panel a shows the evolution of the total $(m,n)$
coarse-grained entropy for an initial $c_{1,1}$ perturbation.  The
thin lines show the raw values, while the thick versions correspond to
a smoothed set of values.  Long- and short-dashed lines represent
test-particle and self-gravitating systems, respectively.  Panel b
shows the time-derivative of the smoothed curves from panel a.  The
solid line corresponds to the test-particle evolution, while the
dot-dashed line reflects the changes in the self-gravitating behavior.
Panel c contains the thermodynamic estimate of $\dot{S}$ based on
Equation~\ref{screate}.  The line styles in this panel are the same as
in panel b.
\label{dsdt}}
\end{figure}

The entropy time-derivative can be manipulated further to cast it as
two terms; one that represents a divergence of a flux and another that
represents entropy creation.  The result is that the entropy creation
term takes the form,
\begin{equation}\label{screate}
\sigma \equiv \frac{\Lambda}{2\avgvv^2}(\avgvvv - 3 \avgvv \avgv +
2 \avgv^3) \frac{\dif \avgvv}{\dif \chi}.
\end{equation}
For an equilibrium situation, the spatial derivative of \avgvv\ is
zero, guaranteeing that the state is an entropy maximum.  Given this
expression, the creation portion of the entropy time-derivative can be
estimated by integrating Equation~\ref{screate} over all space.  The
average velocity values required can be calculated straightforwardly
from first- and second-order coefficient values. As
shown in Figure~\ref{dsdt}, this thermodynamic calculation produces
$\dot{S}$ values that are comparable to those determined from the
coarse-grained values discussed above.

\section{Energy Distributions}\label{nofe}

Entropy evolution is only marginally affected by violent relaxation.
Time-independent mode strengths provide a way to quantify the
incompleteness of a relaxation, and those modes are different between
test-particle and self-gravitating systems.  In this section, we
discuss how energy distributions $n(\beta E)$, the number of particles
with a given energy, can isolate the impact of violent relaxation.
Test particle systems cannot show any $n(\beta E)$ evolution as their
potentials are fixed.  On the other hand, the potential oscillations
that accompany violent relaxation allow mass to change its energy.
The total change to $n(\beta E)$ can be broken into contributions from
different $c_{m,n}$ as follows.

For a first-order perturbation like that in Equation~\ref{pertf}, we
define an energy-distribution perturbation through,
\begin{equation}
\int \int f_{mn,1} \, \dif \chi \dif \varpi = \int n_{mn,1} \, \dif
(\beta E),
\end{equation}
where the integrals run over all possible values.  As usual, we
understand $n_{mn,1}$ by re-writing the left-hand side of this
equation in terms of an energy integral so that the integrands can be
equated.  With the decomposition in Equation~\ref{ord1}, this leads to
\begin{equation}\label{nep1}
n_{mn,1}(\beta E) = c_{m,n} e^{-\beta E} \int_{q1}^{q2} \frac{H_m
\left[ \sqrt{\beta E - \beta \Phi(q)} \right] P_n(q) e^{\beta
\Phi(q)}} {\sqrt{\beta E - \beta \Phi(q)}} \, \dif q.
\end{equation}

\begin{figure}
\scalebox{0.5}{
\includegraphics{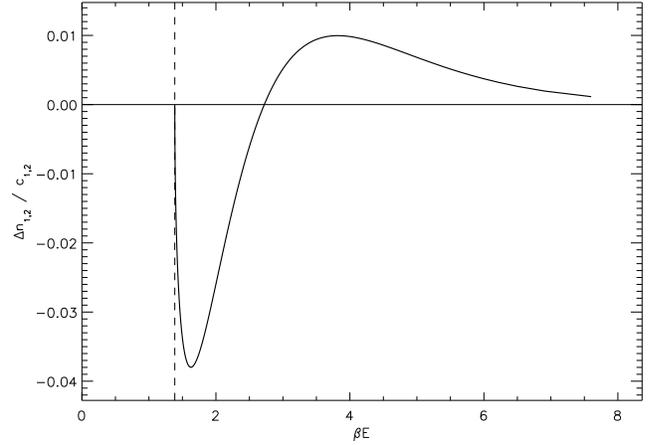}}
\caption{Based on Equation~\ref{nep1}, the impact of a single
perturbing coefficient on the energy distribution can be calculated.
Shown here is the shape of the change induced by a first-order
$c_{1,2}$ perturbation.  Note that depending on the sign of the
perturbing coefficient, the loss/gain in particles can be made to
occur for either higher or lower energies.
\label{nepert}}
\end{figure}

Here, $q=\tanh{\chi}$, $q1=\tanh{\chi_{\rm max}}$,
$q2=-\tanh{\chi_{\rm max}}$, and $\chi_{\rm max}$ is the turning point
location found by solving $\beta E = \beta \Phi(\chi_{\rm max})$.
Equation~\ref{nep1} gives us the change from an equilibrium $n(\beta
E)$ that arises from any given $(m,n)$ coefficient.  As an example,
Figure~\ref{nepert} shows the change in the energy distribution due to
a first-order $m=1$, $n=2$ perturbation.  This perturbation was chosen
because it is non-energetic and allows for substantial simplification
in Equation~\ref{nep1}.  For $m=1$ perturbations, the Hermite term and
the radical in the denominator of Equation~\ref{nep1} cancel.  With
this, it is straightforward to show that a perturbation like $m=1$,
$n=1$ will lead to no change in $n(\beta E)$.  Changes to $n(\beta E)$
due to energetic perturbations are more involved, but their overall
behavior is similar to that shown in Figure~\ref{nepert}.  Depending
on the sign of the perturbing coefficient, particles can be shifted
towards lower or higher energies.  These types of calculations coupled
with coefficient evolutions can be used to determine how the complete
energy distribution perturbation,
\begin{equation}
n_1(\beta E) = \sum_{m,n} n_{mn,1} (\beta E),
\end{equation}
evolves.

\begin{figure}
\scalebox{0.5}{
\includegraphics{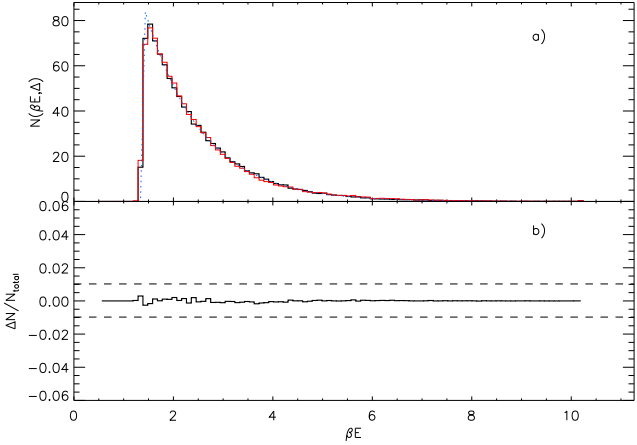}}
\caption{Approximating changes to $n(\beta E)$ using histograms of
particle numbers in an $N$-body simulation of a system with an initial
$c_{1,1}$ perturbation.  Panel a shows an initial state histogram
(thick line), a final steady-state histogram (thin line), and a curve
showing how a separable equilibrium histogram would behave (thin
dotted line) given the same bin widths.  Panel b shows the difference
between the two histograms (solid line) along with average
error-in-the-mean values determined from the ensembles (dashed lines). 
\label{c11nofe}}
\end{figure}

From self-gravitating $N$-body simulations, we can verify these
relationships by approximating $n(\beta E)$ distributions with
histograms of the numbers of particles in finite width energy bins.
For an initial $c_{1,1}$ perturbation, Figure~\ref{c11nofe}
illustrates that 1) the distribution is initially the same as the
separable equilibrium (as expected based on Equation~\ref{nep1}) and
2) there is no net change to the energy distribution once the system
has reached a steady-state.  It is important to realize that $n(\beta
E)$ changes during the evolution, but eventually settles back to
equilibrium.  Any (odd-$m$, odd-$n$) perturbation will behave
similarly.

\begin{figure}
\scalebox{0.5}{
\includegraphics{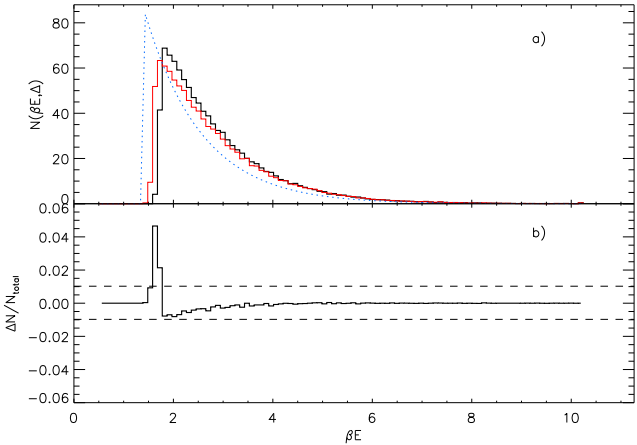}}
\caption{Approximating changes to $n(\beta E)$ using histograms of
particle numbers in an $N$-body simulation of a system with an initial
$c_{0,2}$ perturbation.  The panels and line styles are the same as
those in Figure~\ref{c11nofe}.  We argue that the shift in the
distribution shape is real as the differences are bigger than or
comparable to the statistical fluctuations.
\label{c02nofe}}
\end{figure}

\begin{figure}
\scalebox{0.5}{
\includegraphics{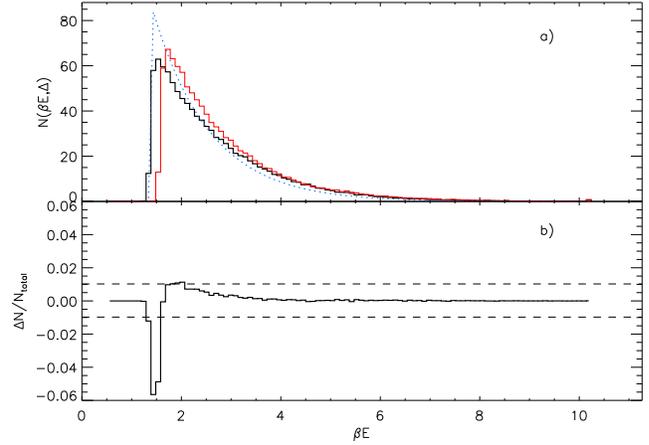}}
\caption{Approximating changes to $n(\beta E)$ using histograms of
particle numbers in an $N$-body simulation of a system with an initial
$c_{2,0}$ perturbation.  The panels and line styles are the same as
those in Figure~\ref{c11nofe}.
\label{c20nofe}}
\end{figure}

For contrast, Figures~\ref{c02nofe} and \ref{c20nofe} show how the
energetic perturbations $c_{0,2}$ and $c_{2,0}$ alter the energy
distribution and lead to permanent distribution changes, respectively.
We argue that these changes are real as the shift in the peak shape
seen involves changes that are larger than, or at least comparable to,
statistical uncertainties in bin occupations.  At no point do the
distributions follow the equilibrium behavior.

The $n(\beta E)$ distribution for a $c_{1,1}$ perturbation oscillates
between deviations like those in Figures~\ref{c02nofe} and
\ref{c20nofe} at various points in its early evolution, before
reaching a steady-state.  These changes to $n(\beta E)$ are easy to
visually inspect, but we want to quantify the impact of violent
relaxation.  To do so, we use the common chi-squared test for
differences between binned distributions \citep{petal94}.  We
calculate the $\chi^2$ value between the initial and subsequent
distributions as a function of time.  At each output time, the
difference between the distributions is squared and then normalized by
the sum of the distributions' values in each bin.

Integrating $\chi^2 (t)$ over the evolution gives us a measure of the
impact of violent relaxation.  For a test-particle system, there is
never any change to $n(\beta E)$, $\chi^2 (t)=0$, and integrating over
time gives zero.  For self-gravitating perturbations, $\chi^2$
asymptotes to a constant value.  A $c_{1,1}$ perturbation has
$\chi^2(t=\infty)=0$, as it returns to equilibrium.  However,
$c_{0,2}$ and $c_{2,0}$ perturbations have non-zero values of
$\chi^2(t=\infty)$.  We calculate the violent relaxation measure as,
\begin{equation}
\mathcal{V}=\int_0^{\infty} [\chi^2(t) - \chi^2(\infty)] \; \dif t.
\end{equation}
$\mathcal{V}$ values are negative if $\chi^2(t)$ tends to be lower
than $\chi^2(\infty)$ during relaxation.  In the systems investigated
here, $\chi^2 (t)$ is an oscillating function.  A negative
$\mathcal{V}$ value indicates that a system spends more time closer to
its original energy distribution when compared to a system with a
positive $\mathcal{V}$ value.  For perturbations like the ones shown
in Figures~\ref{c11nofe}, \ref{c02nofe}, and \ref{c20nofe}, the
$c_{0,2}$ case produces a negative $\mathcal{V}$ value.  With the same
perturbation strength, $c_{1,1}$ perturbations produce $|\mathcal{V}|$
values that are many times larger than those resulting from $c_{0,2}$
and $c_{2,0}$ perturbations.

Given that a system with a $c_{0,2}$ perturbation is closer to a
time-independent state than one with a $c_{2,0}$ perturbation, which
is still closer than one with a $c_{1,1}$ perturbation, we suggest the
following interpretation of self-gravitating $\mathcal{V}$ values.
Negative values reflect a system that undergoes relatively little
relaxation.  The most positive values, for a given perturbation
strength, indicate that systems will return to their original
separable equilibrium after undergoing substantial relaxation.

\section{Summary}\label{sum}

Collisionless one-dimensional gravitating systems are taken as
testbeds for analyzing relaxation processes.  Unlike
three-dimensional situations, the one-dimensional models investigated
here possess separable-solution equilibria with Boltzmann
form that we perturb.  Using second-order perturbation theory, we
investigate relaxation of these systems in terms of entropy
production.  Coefficient dynamics simulations allow us to track fine-
and coarse-grained entropy behavior as perturbed systems settle to
steady states.

We have presented two specific routes for calculating coarse-grained
entropy.  One is based on the more traditional ``binning'' of phase
space which looks at how a distribution function changes over
finite-sized regions of phase space, $\Delta \chi \Delta \varpi$.  The
more appealing definition coarse-grains in a coefficient space formed
by decomposing distributions as series of Hermite-Legendre function
products.  Ignoring small-scale structure by including only low-order
coefficients (small $m$ and $n$ values) provides us with a simple
conceptual picture, when combined with coefficient dynamics.  General
perturbations have time-dependent and time-independent components.
Time-dependent perturbations involve oscillating coefficients that
decay as a wave-like pattern expands to larger $m$ and $n$ values.
These waves represent phase mixing and as the coefficients inside a
coarse-graining box decrease, the entropy increases.  On the other
hand, any time-independent component that is present leaves an imprint
on the coarse-grained entropy.  These time-independent modes
essentially limit the entropy that can be gained by the system.  Their
presence guarantees that the separable equilibrium cannot be
reached through either phase mixing or violent relaxation routes.

In the terminology of \citet{tetal86}, using Maxwell-Boltzmann (or
Lynden-Bell) entropy as an H-function demonstrates that their
collisionless analogue to the H-theorem holds.  Unlike the entropy
behavior determined by the collisional H-theorem, relaxation in these
systems cannot be guaranteed to monotonically increase entropy.
However, terms strongly impacted by phase mixing dominate overall
changes in entropy, leading to increases.

One point of interest is that phase mixing appears to have a much
larger impact on entropy than does violent relaxation.  Our initial
expectation was that self-gravitating systems should show faster
and/or larger entropy changes as a result of the additional relaxation
mechanism.  However, test-particle simulations show roughly the same
increases over the same time-scales as those corresponding to
self-gravitating systems.  The impact of violent relaxation can be
quantified according to how a system's energy distribution changes
during an evolution.

\appendix
\section{Test-particle Coefficient Dynamics}\label{testdem}

This appendix demonstrates that
\begin{equation}\label{test1}
\frac{\partial}{\partial \tau} \sum_{m,n \ge 0} \frac{2^m m!}{2n+1} 
\left[c_{m,n}^{\rm test}\right]^2 = 0.
\end{equation}
To do this we will use the test-particle coefficient dynamics
equations that link diagonal nearest-neighbor values to
time-derivatives in the following way,
\begin{eqnarray}
\lefteqn{\dot{c}_{j,k}^{\rm test} = \frac{k(k-1)}{2(2n-1)} 
c_{j-1,k-1}^{\rm test} - \frac{(k+1)(k+2)}{2(2n+3)} 
c_{j-1,k+1}^{\rm test} +} & & \nonumber \\
& & \frac{(j+1)k(k+1)}{2n-1} c_{j+1,k-1}^{\rm test} -
\frac{(j+1)k(k+1)}{2n+3} c_{j+1,k+1}^{\rm test}.
\end{eqnarray}
For concreteness, we isolate two nearest-neighbor points, $(j,k)$ and
$(j+1,k+1)$.  Expanding the $(m=j,n=k)$ term in Equation~\ref{test1}
using the coefficient dynamics equations leads to a link to the
$(j+1,k+1)$ coefficient,
\begin{equation}
\frac{-2^{j+1} (j+1)! k(k+1)}{(2n+1)(2n+3)}c_{j,k}^{\rm test}
c_{j+1,k+1}^{\rm test}.
\end{equation}
Writing the $(m=j+1,n=k+1)$ term in Equation~\ref{test1} and using the
coefficient dynamics equations produces a link to the $(j,k)$
coefficient,
\begin{equation}
\frac{2^{j+1} (j+1)! k(k+1)}{(2n+1)(2n+3)}c_{j,k}^{\rm test}
c_{j+1,k+1}^{\rm test}.
\end{equation}
Similar arguments can be made for the $(m=j-1,n=k+1)$,
$(m=j-1,n=k-1)$, and $(m=j+1,n=k-1)$ terms.  The quantity being summed
in Equation~\ref{test1} is transferred between terms, but is not
created or destroyed.  Summing over all possible values of $m$ and $n$
guarantees that Equation~\ref{test1} is true.

\section{Second-order $(x,v)$ Coarse-grained Entropy}\label{xvcgs2}

\begin{figure}
\scalebox{0.5}{
\includegraphics{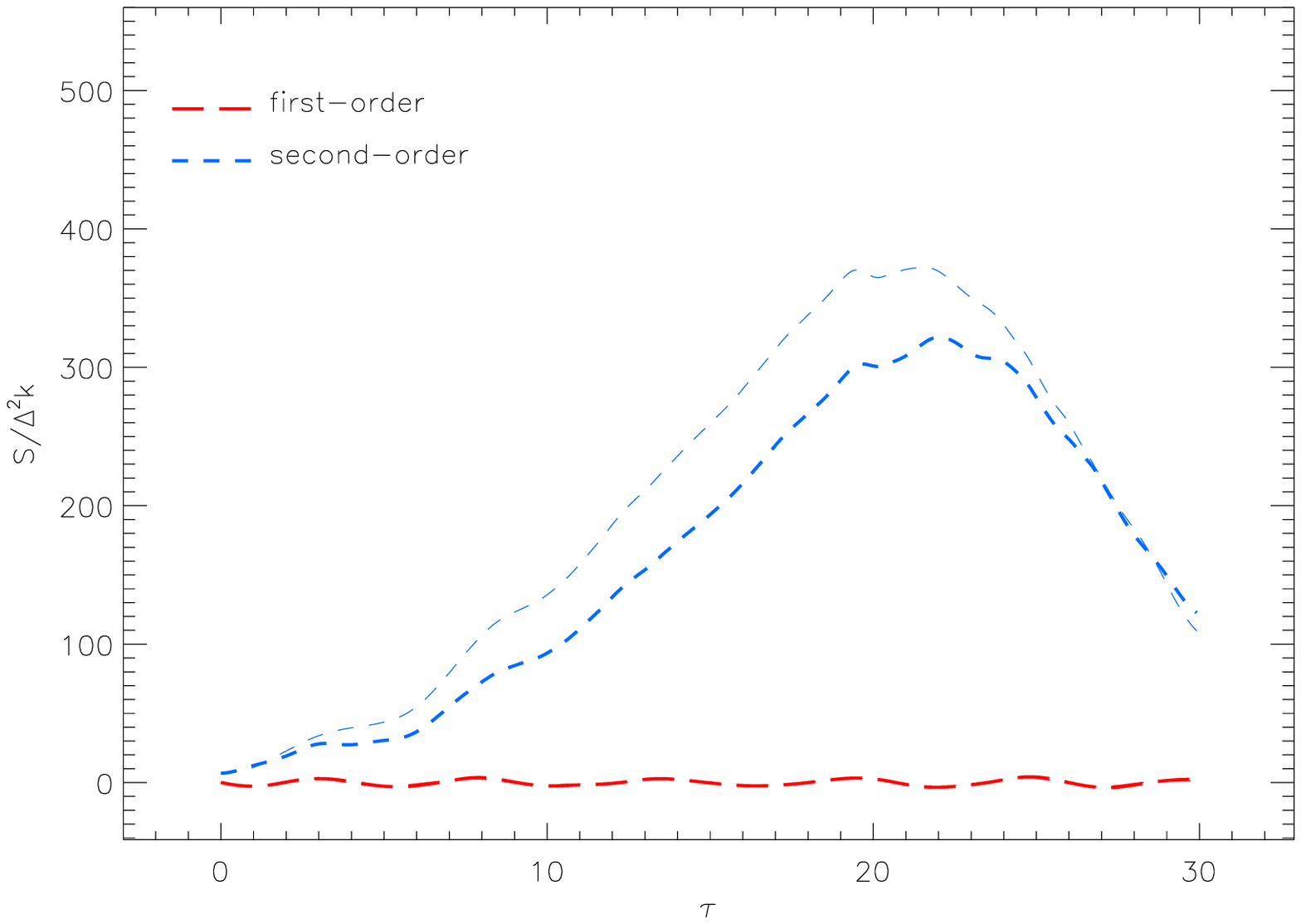}}
\caption{Evolution of the $(x,v)$ coarse-grained entropy for a system
with an initial $c_{1,1}$ perturbation.  Thin lines show the
behavior for a test particle case, thick lines correspond to the
self-gravitating case.  As with the $(m,n)$ coarse-graining,
first-order coarse-grained entropy is essentially constant.  Changes
are appreciable at second-order only.
\label{xvcgsevo}}
\end{figure}

As with the first-order coarse-grained entropy expressions
(Equations~\ref{focgs1} and \ref{focgs2}), the $(x,v)$ coarse-graining
produces a second-order entropy expression that is much more
complicated than the analogous $(m,n)$ coarse-graining version.  From
Equation~\ref{cgflnf}, the coarse-graining corrections to the second
order entropy are,
\begin{eqnarray}
\lefteqn{\iint \gamma_{xv,2} (1+\ln{f_0}) \, \dif \chi \dif \varpi +}
& & \nonumber \\
& & \frac{1}{2} \iint \frac{\gamma_{xv,1}^2}{f_0} \, \dif \chi \dif
\varpi + \iint \frac{\gamma_{xv,1} f_1}{f_0} \, \dif \chi \dif
\varpi.
\end{eqnarray}
The first term is analogous to the first-order correction term
(Equation~\ref{focgs1}), with $c_{0,n}$ replaced by $d_{0,n}$.  Since
$\gamma_{xv}$ terms involve the small quantity $\Delta^2$, the second
term should be much smaller than the first and third, and we ignore it
here.  Finally, the third term can be shown to be,
\begin{eqnarray}
\lefteqn{\iint \frac{\gamma_{xv,1} f_1}{f_0} \, \dif \chi \dif 
\varpi =} & & \nonumber \\
& & \pi \frac{\Delta^2}{6} \sum_{m,n \atop m\ne n=0} \left[
\frac{2^{m+2}(m+2)!}{2n+1} c_{m+2,n}c_{m,n} + \right. \nonumber \\
& & \frac{2^m m!}{2n+5} (A_n + C_n Q_0^{(2,n)}) c_{m,n+2}c_{m,n} + 
\nonumber \\
& & \frac{2^m m!}{2n+1} (B_n + C_n Q_2^{(2,n)}) c_{m,n}^2 + 
\nonumber \\
& & \left. \frac{2^m m!}{2n-3} C_n Q_4^{(2,n)} c_{m,n-2}c_{m,n}
\right].
\end{eqnarray}
The complexity of the terms in this expression makes a simple
interpretation of the coarse-grained entropy time-behavior difficult.
However, numerically following the coefficient behavior allows us to
calculate this quantity during an evolution.  The results are shown in
Figure~\ref{xvcgsevo}.

\end{document}